\documentclass[12pt]{iopart}
\usepackage{epsfig}

\newcommand{\be}{\begin{equation}}
\newcommand{\ee}{\end{equation}}
\newcommand{\bea}{\begin{eqnarray}}
\newcommand{\eea}{\end{eqnarray}}
\newcommand{\bfig}{\begin{figure}[ht]}
\newcommand{\efig}{\end{figure}}

\begin{document}
\title{Entanglement of qubits via a nonlinear resonator}
\author{M Kurpas, J Dajka and E Zipper}
 
\address{Institute of Physics, University of Silesia, Ul. Uniwersytecka 4, 40-007 Katowice, Poland}
\ead{mkurpas@us.edu.pl}

\date{\today}

\begin{abstract}

Coherent coupling of two qubits mediated by a nonlinear resonator is studied. It is shown that the
amount of entanglement accessible in the evolution depends both on the
strength of nonlinearity in the Hamiltonian of the resonator and on the initial preparation of the system.
The created entanglement survives in the presence of decoherence.
\end{abstract}
\maketitle

\section{Introduction}
Developments in quantum information science rely critically on
entanglement - quantum correlations between two or more physical systems e.g. between two qubits.
There are various ways of creating entangled states of two qubits. The most natural is to place the qubits close to each other so that they can interact directly by local interactions i.e. a mutual inductance or capacitance. There are spectacular experiments performed for solid state qubits using this type of interaction \cite{mooij,pashkin,berkley,izmalkov,ploeg}. However the important stetp towards implementing quantum computation is a controllable coupling of qubits.

The entanglement of distant qubits can be reached by swapping of entanglement \cite{moehring,kok,my} or by some 'interaction--carrying' medium: a bus. In the following we discuss the latter case. One of natural candidates for such a bus are photons that are highly coherent. The problem of interaction of qubits with a quantized monochromatic electromagnetic field in a quantum cavity has been discussed in several papers \cite{blais1,migliore,zag,majer,sill}. The field mediating the entanglement is usually described by a linear resonator \cite{blais1,migliore,zag,paraoanu,li}. However, it has been shown \cite{zag} that strong entanglement of qubits in linear coupling regime is possible only for specific initial states. In this paper we want to check wheather a nonlinear radiation improves this situation.

Nonlinearity of the superconducting circuit resonator operating at microwave frequencies results in highly non--classical properties of the electromagnetic radiation generated by the resonator \cite{n1}. Such circuits coupled to the qubits has recently been intensively studied \cite{zhou} since they do provide a natural measuring device \cite{siddiqi,n2}. On the other hand, the problem of nonlinear oscillator mediating the interaction between qubits is still open for theoretical investigations. A pair of qubits coupled via a driven Duffing oscillator has been studied in \cite{n3} with an emphasis given on the (semi)classical limit of the oscillator. In some coupling schemes \cite{blais,nisk1,niskanen} an intermediate qubit with one or more Josephson junctions plays a role of a coupler. As the junction is strongly nonlinear, the coupler is treated as a nonlinear resonator. 

A nonlinear character of coupling has been also broadly exploited in quantum optics: there are many proposals where nonlinear coupling effects play an essential role in generation of entanglement of photons \cite{gerry,leonski,yi,zhang} and in construction of quantum gates \cite{ottaviani} or teleportation protocols \cite{vitali}.

In this paper we assume two qubits strongly coupled to a resonator and discuss a coherent, non-local coupling between qubits via a linear and nonlinear resonator. The goal is to determine conditions which allow
to strongly entangle the qubits with each other with the disentangled resonator. Thus the careful study of properties of the quantum bus is needed in order to fulfiil this task. We investigate the influence of the strength of nonlinearity on coherent coupling of qubits and we show that it is highly nontrivial. On the basis of obtained results we propose a scheme of a tunable entangling gate.

The results are applicable to entanglement through optical and microwave nonlinear radiation under the assumption that the qubit-field interaction is sufficiently weak to be described in the rotating wave approximation \cite{gardiner}.

In the first part of the paper we assume that all effects of dissipation and decoherence are negligible over studied time scales. We briefly discuss the effect of dissipation in the last section of the paper. To quantify the entanglement we calculate the negativity $N$
\cite{neg} which is a natural measure of entanglement. 
Our discussion is general enough to not depend on the specific architecture. However, a candidate for
possible implementation is the flux or charge qubit interacting with microwave radiation.

In Section \ref{model} we present the model of the system, in Sec. \ref{gain} and \ref{suppression} we trace the evolution of the originally disentangled state with a {\it single excitation}. We consider two initial states which will be shown to result in qualitatively different behaviour of the qubit--qubit entanglement:
nonlinearity--assisted entanglement gain or entanglement suppression. In Section \ref{many_excitations} we investigate the system with {\it two} and {\it three excitations}, in Sec. \ref{qq_matrix} we verify the results for the negativity by presenting the qubit-qubit density matrix. Finally in Sec. \ref{decoh} we shortly study the influence of decoherence on the qubits entanglement. Discussion and conclusions are given in Sec. \ref{disc}.

\section{The Model}
\label{model}
The system under considerations consists of two qubits $Q_i$, $i=1,2$  and
a monochromatic resonator $R$. We assume that both qubits are
coupled to the resonator simultaneously. To simplify the notation we apply units such that $\hbar=1$.
In the absence of direct qubit-qubit interaction the Hamiltonian of such a
system takes the form
\be
H = H_{Q_{1}}+ H_{Q_{2}}+ H_{R}+H_{Q_{1}R} + H_{Q_{2}R},
\ee
the qubit Hamiltonian is
\be
 H_{Q_{i}} =\frac{  \Omega_{i}}{2} \sigma _z,
\ee
The interaction terms $H_{Q_iR}$  read
\be
H_{Q_iR}=  - \frac{\gamma}{2}\left( a\sigma^{+}+a^{\dagger }\sigma^{-}\right) \label{hqr}
\ee
where $\sigma^+=\sigma _x+i\sigma_y$, $\sigma^-=(\sigma^+)^\dagger$ and
$\sigma _z$ are the Pauli matrices and $\gamma$ is the qubit-field coupling constant depending on the specific architecture.
The resonator is  described by a nonlinear bosonic oscillator of
the Hamiltonian
\be
H_R=\omega_R \left( a^{\dagger }a+\frac {1}{2}\right)+ V_{R},
\label{hr}
\ee
where the term $ V_{R}$ describes the nonlinearity.
We shall  discuss  two different forms of the nonlinearity. The first is a
polynomial one
\bea
 V^1_{R} &= \alpha \left( a ^2  + \left(a^{\dagger}\right)^2 \right)
\label{aap}
\eea
which is known to be related to the celebrated squeezed states
\cite{vourdas1,perina}.
The second one, of central interest for our considerations, is the
cosine--like nonlinearity
\bea
 V^2_{R} &= \alpha \cos(a + a^{\dagger})
\eea
The cosine term introduces nonlinearities of all orders and in that sense
the cosine is 'more nonlinear' than any polynomial. This type of
nonlinearity is present e.g. in the Hamiltonian of the microwaves generated
by a SQUID \cite{mooij,niskanen,ve}. For such a circuit the parameter $\alpha$ can be changed by introducing into the resonator an adjustable weak link loop with two Josephson junctions. The Josephson tunneling energy controllable can by an external flux $\phi_c$; in this case $\alpha=2E_J^0\cos(\pi\phi_c/\phi_0)$ \cite{ve,makh}. Varying $\phi_c$ we can change $\alpha$ between $2E_J^0$ and $0$. A tunable anharmonic $LC$ circuit is also important as it introduces non--uniform level spacing reducing leakage to higher states \cite{blais}.
In the limit of $V_R= 0$ one arrives at the Hamiltonian of the
Jaynes--Cummings type \cite{cos2} for which exact solutions are known
\cite{zag}.

The state vector of the system at $t=0$ is a tensor product of three
constituent states
\be
 \vert \psi _{Q_{1}Q_{2}R}(t=0) \rangle =\vert \psi_{Q_{1}}\rangle \otimes
 \vert \psi_{Q_{2}} \rangle \otimes \vert\psi_{R} \rangle
 \ee
This choice is certainly justified for weakly interacting systems.
The unitary evolution of the initially factorisable state leads in general
to the entangled tripartite state. By taking the trace over the
$R$ states one obtains the reduced density operator $\rho _{Q_1Q_2}$ which we then use to calculate the negativity  \cite{neg}
\be
N(\rho_{Q_1Q_2})=\max(0,-\sum_i \lambda_i),
\ee
where $\lambda_i$ are negative
eigenvalues of the partially transposed \cite{peres} density matrix of the two
qubits. $N$ is 0 for separable states and reaches its maximal value $N=1/2$ for maximally entangled states. \\
We look for the conditions under which the qubits get very strongly entangled.

\section{Entanglement and nonlinearity} 
\label{ent_nonlinear}

In the following we study the influence of the nonlinear term in the
bosonic Hamiltonian (\ref{hr}) on the quantitative and qualitative properties of
entanglement of the qubit--qubit system.
We consider below only the cosine nonlinearity as we have found that the influence of the nonlinearity (\ref{aap}) is weaker and not qualitatively different.
We compare the results with those obtained for the linear single mode
resonator, which despite its simplicity is a natural reference system. For numerical analysis  we truncated the photonic space to $M=40$ for which $Tr( \vert \psi _{Q_{1}Q_{2}R}(t) \rangle\langle \psi _{Q_{1}Q_{2}R}(t)\vert\simeq1)$. 

\subsection{Entanglement gain}
\label{gain}
The first of the considered initial states is
\be \vert \psi _{Q_{1}Q_{2}R}(t=0) \rangle = \vert eg0\rangle
\label{eg0}
\ee
Here $e$ and $g$ stand for the excited and ground state of
one qubit respectively.
The state (\ref{eg0}) describes the first qubit in an excited state, the second one in the ground state
and there are no photons inside the resonator.

Let us start with an analysis of the system in the absence of the nonlinear
term i.e. $V_R=0$.
In this case one can perform exact analytical calculations of the wave
function and trace the behavior of the system.
The wave function of the system at the time $t$ is \cite{zag}
\bea
\label{eg0_t}
\vert \psi _{Q_{1}Q_{2} R}(t) \rangle &=& \frac{1}{2}
\left(1+\cos(\tilde{\gamma}t) \right) \vert eg0\rangle  \rangle   \\ &-&\frac{1}{2} \left(1-\cos(\tilde{\gamma}t)\right) \vert ge0\rangle +  \frac{i}{\sqrt2} \sin(\tilde{\gamma}t) \vert gg1\rangle, \nonumber
\eea
where $\tilde{\gamma}=\sqrt{2}\gamma$.
The time evolution goes in the subspace spanned by the states $\vert eg0\rangle
,\vert ge0\rangle, \vert gg1 \rangle$.
For $\tilde{\gamma}t=n\pi$, $n=1,2,...$ for which the $R$ subsystem could
be decoupled from the qubits also the coefficient at $ \vert ge0 \rangle$ term becomes
zero and the whole system remains disentangled. It is shown in Fig.\ref{linear} as lines $A$ ($QQ-R$ entanglement) and $B$ ($QQ$ entanglement) at $\omega t \approx 220$.  For other values of $\tilde{\gamma}t$ we are not able to obtain the entanglement of the qubits without the intrusion of the resonator states. This results in weak $QQ$ entanglement with the limiting value $N_{QQ}^{max}\sim 0.1$.
 %When the $QQ-R$ entanglement (the entanglement between the qubits as a one composite and the resonator) is strong, a lot of information about the system is shared between $QQ$ and $R$. At the same time, there is little information shared between the qubits. Tracing out the resonator gives loss of a big part of the system information, thus the qubit-qubit entanglement is weak.
% It is shown that $N_{QQ}$ oscillates with the frequency proportional to $\gamma$ reaching $N_{QQ}^{max}=0.1$.
\bfig
\centering
\includegraphics[width=0.6\textwidth]{fig1.eps}
\caption{(color online) The $QQ-R$ (line A) and $QQ$ (line B) negativity as a function of dimensionless time $\omega t$ for the linear case and for the initial state $\vert eg0\rangle$. Lines $C$ and $D$ depict the $QQ$-$R$ and $QQ$ negativity for the initial state $\vert gg1\rangle$; $\Omega_i=\omega, \gamma=0.01\omega$. Maximal qubit-qubit entanglement is possible when the resonator is not entangled with the qubits (lines $C$ and $D$ at $\omega t = 110$).}
\label{fig1}
\label{linear}
\efig
 %This effect occurs due to very specific relation of the symmetry of the initial state and the Hamiltonian of the full system.
The inclusion of the nonlinear term ($V_R \neq 0$) allows to go beyond this limit. It is shown in Fig.\ref{fig2} that even for
the relatively weak nonlinearity the negativity remains bounded
by nothing but its natural limit i.e. $N_{QQ}^{max}=1/2$.
\\ 
Increasing the strength of the nonlinearity (Fig.\ref{fig3}) the qubits
get permanently entangled for a relatively long time which
increases with increasing $\alpha$. For very strong nonlinearity ($\alpha = 2$) the period is $25000 \omega t$ and the maximal value of $N_{QQ}\approx 0.45$ (not shown).
Thus we see that for the non-symmetric initial state, when the qubits start from
opposite states, the nonlinear resonator can lead, in contrast to the linear one, to the emergence of
strongly entangled $QQ$ states remaining very weakly entangled with  the resonator states. In other words we get the coherent quantum state transfer between the qubits through the quantum bus.
\bfig[h]
\centering
\includegraphics[width=0.6\textwidth]{fig2.eps}
\caption{(color online) The unitary evolution of the qubit-qubit entanglement as
a function of dimensionless time $\omega t$ and the nonlinearity coefficient $\alpha$.
The initial state $\vert eg0 \rangle$. $\Omega_i=\omega, \gamma=0.01\omega$}
\label{fig2}
\efig  
\bfig[h]
\centering
\includegraphics[width=0.6\textwidth]{fig3.eps}
  \caption{(color online) The qubit-qubit negativity for large values of the nonlinearity coefficient $\alpha$.
The initial state $\vert eg0 \rangle$. $\Omega_i=\omega, \gamma=0.01\omega$.}
\label{fig3}
\efig

\subsection{Entanglement suppression}
\label{suppression}
 The situation looks different for the 'symmetric' initial state
\be
\vert \psi _{Q_{1}Q_{2}R}(t=0) \rangle=\vert gg1 \rangle,
\label{gg1}
\ee
when the excitation is placed in the resonator and both qubits are in
the ground state.
The analytical calculations for the linear resonator ($V_R=0$) 
 result in the wave function \cite{zag}
 \bea
 \vert \psi _{Q_{1}Q_{2}R}(t)\rangle&=& \frac{i}{\sqrt{2}}\left[  \sin(\tilde{\gamma}t)
 \vert eg0 \rangle +\sin(\tilde{\gamma}t) \vert ge0 \rangle \right]  \nonumber \\ &&+\cos(\tilde{\gamma}t)\vert gg1 \rangle.
\label{eq13}
 \eea
 We see that at $\tilde{\gamma}t= \pi/2+m\pi$, where $m$ is an integer, we
 get maximally entangled qubit-qubit (Bell) state
 \be
 \vert B_{1}\rangle=\frac 1{\sqrt{2}}(\vert eg\rangle + \vert ge\rangle ).
 \ee
 It is important that the qubits can be in the maximally entangled state only, when their common state space is separable from the resonator space. It is shown in Fig.\ref{fig1} (lines $C$ - the $QQ$-$R$ negativity and $D$ - the $QQ$ negativity) at $\omega t = 110$. As in the case of the initial state $\vert eg0\rangle$, the qubit-qubit negativity $N_{QQ}$ reaches the value 0.1 (line $D$) when the $QQ$-$R$ entanglement is maximal (Fig.\ref{fig1} $\omega t = 50$ line $C$) .

 The inclusion of the nonlinear term spoils the symmetry (see Eq. (\ref{eq13})) and results in suppression of the qubit--qubit entanglement with the negativity limited by $N_{QQ}^{max}<1/2$ depending on the amplitude $\alpha$ (see Fig.\ref{fig4}). 

 \bfig
\centering
\includegraphics[width=0.6\textwidth]{fig4.eps}
\caption{(color online) The influence of nonlinearity strength $\alpha$ of the resonator on the entanglement of the qubits for the initial state (\ref{gg1}). $\Omega_i=\omega, \gamma=0.01\omega$.}
\label{fig4}
\efig

 \section{States with many excitations}
 \label{many_excitations}
Finally, let us proceed to a consideration of the evolution of the originally disentangled states with many excitations. In Figs \ref{fig5} and \ref{fig6} we show the resulting qubit-qubit entanglement for the initial state with two and three excitations respectively
 \bea
 \vert \psi _{Q_{1}Q_{2}R}(t=0)\rangle &=&\vert eg1\rangle  \label{eg1}\\
 \vert \psi _{Q_{1}Q_{2}R}(t=0) \rangle
 &=&\vert eg2\rangle.  
\label{eg2}
 \eea
 If the interaction between the qubits goes via a linear resonator ($\alpha=0$) the negativity $N_{QQ}$ (and hence the $QQ$ entanglement) is very small (solid line). In other words, one is not able to get rid of the resonator degrees of freedom without leakage of the information about the state of the system. The situation changes if the interaction goes via the nonlinear resonator ($\alpha\neq 0$ lines in Figs \ref{fig5} and \ref{fig6}) and we obtain strong $QQ$ entanglement. Thus in this case the influence of the nonlinearity of the "quantum bus" is favorable for the creation of the entanglement of the qubits.
\\
 If we start from 
 \be
 \vert \psi _{Q_{1}Q_{2}R}(0)\rangle = \vert ee0\rangle
 \label{ee0}
 \ee
 the interaction via the linear resonator does not entangle the qubits ($N_{QQ}=0$).
 It is in agreement with exact analytical calculations \cite{zag}. We have also checked it by calculating the concurrence (not shown) which is another entanglement measure. Surprisingly, switching "on" the nonlinearity does not improve the situation and the qubits remain disentangled.

For the initial state
 \be
 \vert \psi(0)\rangle = \vert gg2\rangle
\ee
for the interaction via the linear resonator we get the $QQ$ entanglement with $N^{max}_{QQ}\approx 0.18$. The presence of nonlinearity leads, in general, to a decrease of $N^{max}_{QQ}$ but for some specific $\alpha$ ($\alpha \approx 0.7$) we have found a small increase of the negativity (not shown).
  
Summarizing, the influence of nonlinearity is favorable for the initial states with one qubit in the excited state.
Then the coupling via the quantum bus leads to strong entanglement and a coherent state transfer between the qubits.
\bfig
\centering
\includegraphics[width=0.6\textwidth]{fig5.eps}
\caption{(color online) The influence of the nonlinearity strength $\alpha$ on the qubit-qubit entanglement for the initial state $\vert eg1\rangle$, $\Omega_i=\omega, \gamma=0.01\omega$.}
\label{fig5}
\efig
\bfig
\centering
\includegraphics[width=0.6\textwidth]{fig6.eps}
\caption{(color online) The influence of the nonlinearity strength $\alpha$ on the qubit-qubit entanglement for the initial state $\vert eg2\rangle$, $\Omega_i=\omega, \gamma=0.01\omega$.}
\label{fig6}
\efig

\section{Qubit-qubit density matrix}
\label{qq_matrix}
\bfig
\centering
\includegraphics[width=0.6\textwidth]{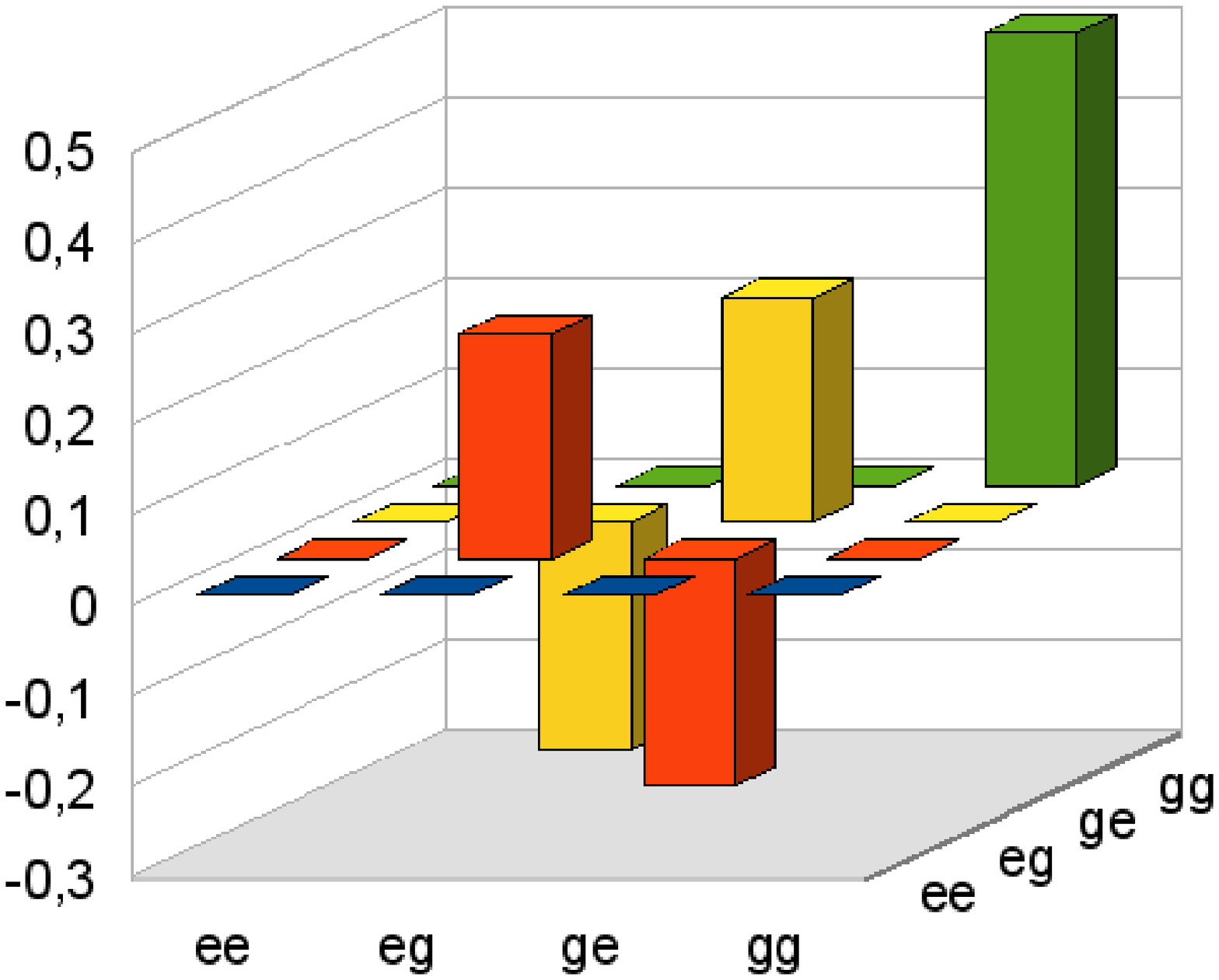}
\caption{(color online) The real part of the qubit-qubit density matrix for the case of the linear resonator and the initial state (\ref{eg0}). The matrix is taken at $\omega t=111$ for which $N_{QQ}=0.104$ (solid line in Fig.\ref{fig2}). The imaginary elements of the matrix are of the order of $10^{-14}$.}
\label{fig7}
\efig

\bfig
\centering
\includegraphics[width=0.6\textwidth]{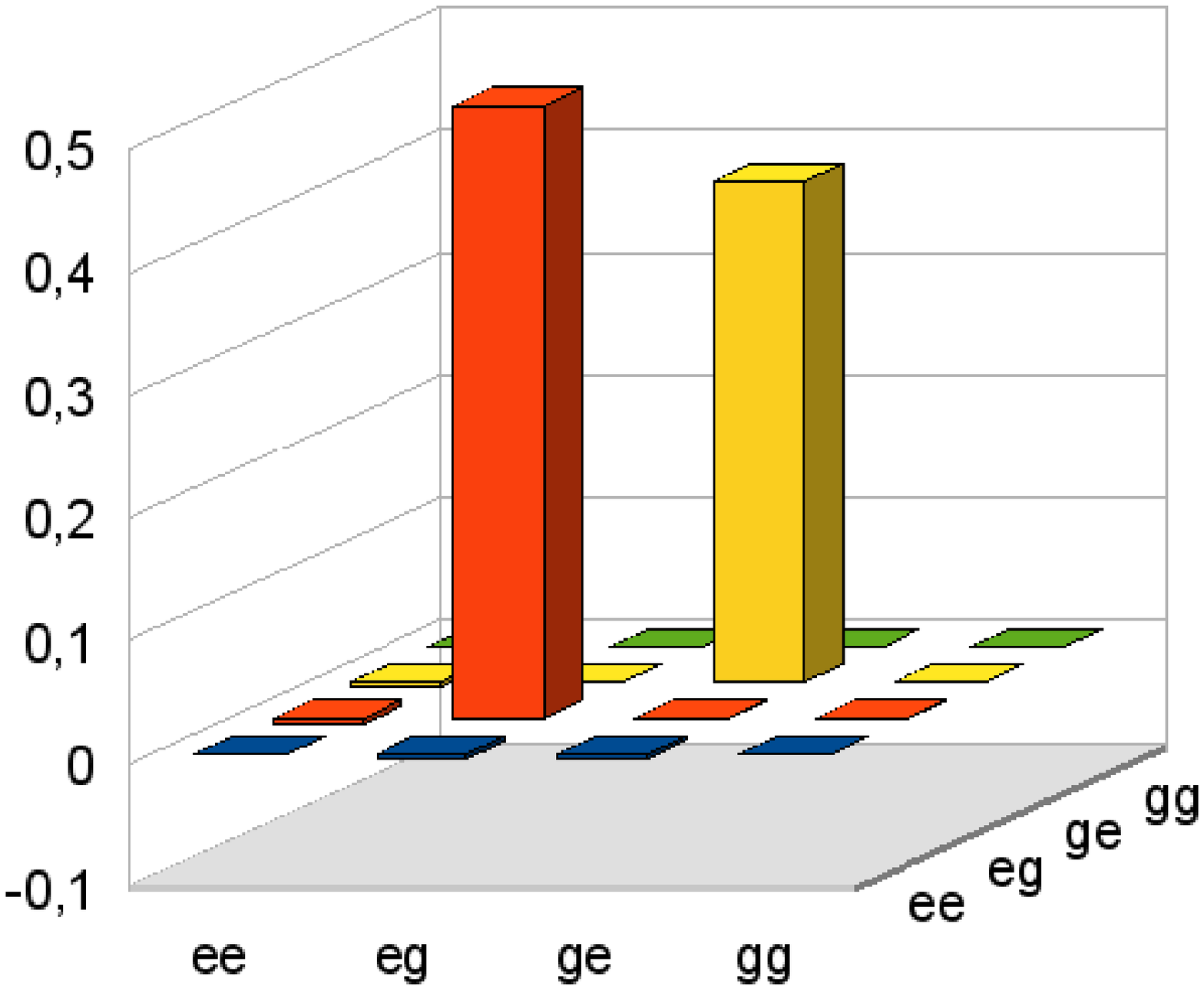}
\caption{(color online) The real part of the qubit-qubit density matrix for the case of the nonlinear resonator ($\alpha=0.0035$) and the initial state (\ref{eg0}). The values are taken at $\omega t=435$, $N_{QQ}=0.492$ (dashed line in Fig.\ref{fig2}) . }
\label{fig8}
\efig
\bfig
\centering
\includegraphics[width=0.6\textwidth]{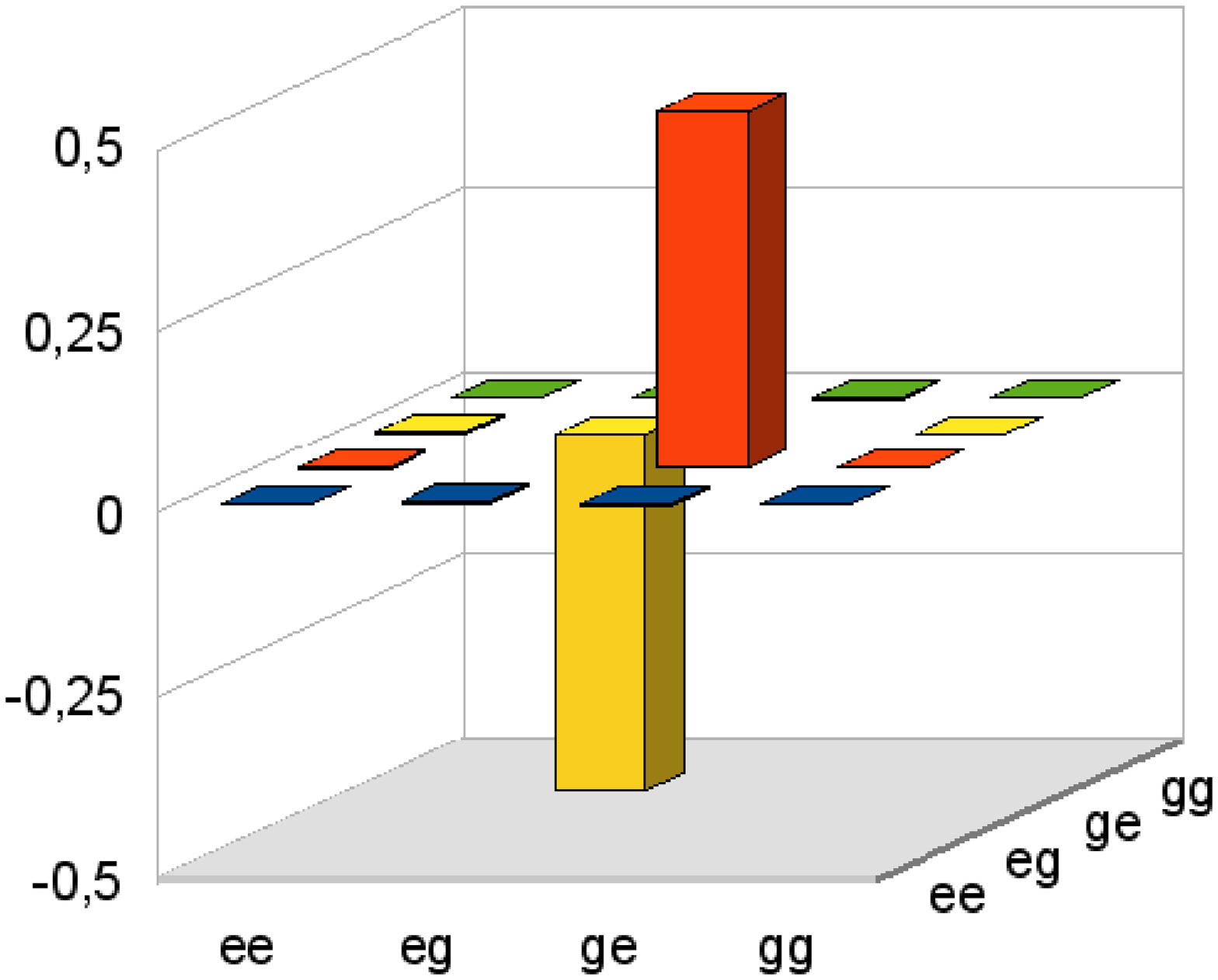}
\caption{(color online) The imaginary part of the qubit-qubit density matrix for the case of the nonlinear resonator ($\alpha=0.0035$) and the initial state (\ref{eg0}). The values are taken at $\omega t=435$, $N_{QQ}=0.492$ (see dashed line in Fig.\ref{fig2}). }
\label{fig9}
\efig
One of the best demonstrations of quantum entanglement is the quantum state tomography which yields a density matrix of coherently coupled qubits \cite{stef}. Such experiments for qubits interacting via a resonant cavity are in progress \cite{sill}.
As an example we present results of our calculations of the qubit-qubit density matrix for the case discussed in Sec.\ref{gain}.

In Fig.\ref{fig7} we show the real part of the density matrix for the qubits interacting via the linear resonator. The matrix elements are taken at $\omega t=111$ for which $N_{QQ}=0.104$ in Fig.\ref{fig2}. Looking at the state vector (\ref{eg0_t}), this situation corresponds to equal probability of finding the excitation in the resonator or in one of the qubits. 
In Fig.\ref{fig8} and  Fig.\ref{fig9} we present the real and imaginary part of the $QQ$ density matrix for $\alpha=0.0035$. We see that in this case the non-diagonal matrix elements being the hallmark of entanglement have  much larger values as when the coupling goes via the linear resonator. The values correspond to dashed line in Fig.\ref{fig2} at $\omega t=435$, $N_{QQ}=0.492$. 

\section{The effect of decoherence}
\label{decoh}
Design and construction of quantum devices is always affected by the
influence of environment. In the following we apply the commonly used Markovian approximation
 \cite{gardiner} and model the reduced dynamics of the $QQR$ system  in
terms of master equation generating complete positive dynamics \cite{alicki}.
We take into account the decoherence of both qubits and the resonator.
Following \cite{blais1,my} we assume that the effect of environment can be included
in terms of two independent Lindblad terms:
\begin{eqnarray}
\dot{\rho}(t)=[L_H-\frac{1}{2}L_{D}]\rho(t)
\end{eqnarray}
where the 'conservative part' is given by
\begin{eqnarray}
L_H(\cdot)=-i[H,\cdot]
\end{eqnarray}
whereas the 'Lindblad dissipator'
\begin{eqnarray}
L_D(\cdot)&=&L_R(\cdot)+\sum_{i=1}2 L_{Q_i}(\cdot)\\
L_R(\cdot)&=&A^\dagger A(\cdot)+(\cdot)A^\dagger A-2A(\cdot)A^\dagger\\
L_{Q_i}(\cdot)&=&\Sigma_i^\dagger \Sigma_i(\cdot)+(\cdot)\Sigma_i^\dagger
\Sigma_i-2\Sigma_i(\cdot)\Sigma_i^\dagger
\end{eqnarray}
 is expressed in terms of creation and annihilation operators 'weighted' by
suitable lifetimes  $A=a/\sqrt{T_{R}}$  and
$\Sigma_i=\sigma^-/\sqrt{T_{Q_i}}$, where $T_R=5 \cdot 10^{-5}s$ and $T_{Q_i}=10^{-5}s$ are the resonator and the qubits decoherence times respectively. 
\bfig
\centering
\includegraphics[width=0.6\textwidth]{fig10.eps}
\caption{(color online) The influence of decoherence on the entanglement of the qubits for the initial state (\ref{eg0}). $\Omega_i=\omega, \gamma=0.01\omega, \alpha=0.0035,~T_R=5 \cdot 10^{-5}, T_{Q_i}=10^{-5}s$. }
\label{fig10}
\efig
It is shown in Fig.\ref{fig10} that the effect of such a local damping is purely
qualitative and the constructive role of of the nonlinear term in the
oscillator is not obscured provided that the relaxation times are sufficiently large.

\section{Discussion and conclusions}
\label{disc}
In this paper we have considered entanglement generation in the system consisting of two qubits coupled by a linear or nonlinear resonator. The goal was to test the role of nonlinearity and to find conditions, under which the qubits became strongly entangled. We have considered cosine--like nonlinearity where the discussed effects become most visible. We have found that the influence of the nonlinearity of the resonator, having
its own dynamics, depends significantly on the initial state of the investigated system. \\
We have shown two qualitatively different behaviours.
The first occurs in the system with non-symmetric initial states. The presence of nonlinearity results in strong enhancement of the entanglement almost up to its maximal value. For sufficiently large amplitude of the nonlinear term, the entanglement becomes (quasi) permanent and does not vanish for a long time. This constructive role of nonlinearity is certainly desired for the applications.
For the symmetric initial states the role of the nonlinearity is no more constructive but rather results in suppression of entanglement. 

On the basis of these findings, we propose the following tunable coupling scheme: for the non-symmetric initial state the nonlinear term should be switched {\it on}, for the symmetric one this term should be {\it off} and the interaction goes via the linear resonator. Then in both cases we obtain the desired strong entanglement of qubits and the system acts as a quantum entangling gate. To our knowledge, such a scheme has not been proposed yet. 

The main advantage of this gate is that the strong entanglement of qubits can be reached for many initial states, even for the states with many excitations (with some exceptions).
We have also shown that the coherent coupling of qubits survives in the presence of dissipation assuming decoherence times in agreement with recent experiments \cite{majer,sill}.

\section*{Acknowledgement}
We thank Jun Jing for valuable remarks.
The work  supported by the Polish Ministry of Science and Higher Education
under the grant N 202 131 32/3786, Scientific Network LEPPI Nr 106/E-
345/BWSN-0166/2008 and the scholarship from the UPGOW project co-financed by the European Social Fund.


\section*{References}
\begin{thebibliography}{10}
\bibitem{mooij} Majer J, Paauw F G, ter Haar  A C J, Harmans C J P M , Mooij J E 2005 Phys. Rev. Lett \textbf{94}, 090501
\bibitem{pashkin} Pashkin Yu A, Yamamoto T, Astafiev O, Nakamura Y, Averin D V and Tsai J S 2003 Nature \textbf{421}, 823

\bibitem{berkley} Berkley A J, Xu H, Ramos R C, Gubrud M A, Strauch F W, Johnson P R, Anderson J R, Dragt A J, Lobb C J, Wellstood F C 2003 Science \textbf{300}, 1548 
\bibitem{izmalkov} Izmalkov A, Grajcar M, Il’ichev E, Wagner Th, Meyer H-G, Smirnov A Yu, Amin M H, Maassen van den Brink A and Zagoskin A M 2004 Phys. Rev. Lett. \textbf{93}, 037003 
\bibitem{ploeg} van der Ploeg S H W et al. 2007 Phys. Rev. Lett. \textbf{98}, 057004 
\bibitem{moehring} Moehring D L, Maunz  P, Olmschenk S, Younge K C, Matsukevich D N, Duan L M, Monroe C 2007 Nature {\bf449}, 68
\bibitem{kok} Barrett  S D, Kok P 2005 Phys. Rev. A {\bf71}, 060310 (R)
\bibitem{my} Zipper E, Kurpas M, Dajka J and Ku\'{s} M 2008  J. Phys: Condensed Matter, \textbf{20}, 275219
\bibitem{blais1} Blais A, Huang R S, Wallraff A, Girvin S M, Schoelkopf R J 2004 Phys. Rev. A {\bf 69}, 062320
\bibitem{migliore} Migliore R, Messina A 2005 Phys. Rev. B \textbf{72}, 214508
\bibitem{zag} Smirnov A Yu, Zagoskin A M 2002 cond-mat/0207214
\bibitem{majer} Majer J, Chow J M, Gambetta  J M, Koch J, Johnson B R, Schreier J A, Frunzio L, Schuster D I, Houck A A, Wallraff A, Blais A, Devoret M H,Girvin S M, and Schoelkopf R J  2007 Nature \textbf{449}, 443
\bibitem{sill} Sillanpaa M A, Park J I, Simmonds R W 2007 Nature \textbf{449}, 438
\bibitem{paraoanu} Paraoanu G S 2006 Phys. Rev. B \textbf{74} 140504(R) 
\bibitem{li} Li  J, Chalapat K, Paraoanu G S 2008 Phys. Rev. B \textbf{78}, 064503
\bibitem{n1} Young J F, Nori F 2008 Phys. Rev. Lett. {\bf 101} 253602
\bibitem{zhou} Zhou X, Mizel A 2006 Phys. Rev. Lett. {\bf 97}, 267201 
\bibitem{siddiqi} Siddiqi I, Vijay R, Pierre F, Wilson C M, Frunzio L, Metcalfe M, Rigetti C, Schoelkopf R J, and Devoret M H, Vion D and Esteve D 2005 Phys. Rev. Lett. {\bf 94}, 027005 % JJ nonlinearity - sterowanie nieliniowoscia, wzmacniacz CQED
\bibitem{n2} Picot T, Lupa\c{s}cu A, Saito S, Harmans C J P M, Mooij J E 2008 Phys. Rev. B {\bf 78}, 132508
\bibitem{n3} Buri\'{c} N 2009 Phys. Rev. A {\bf 79} 022101
\bibitem{blais} Blais A, Maassen van den Brink A, Zagoskin A M 2003 Phys. Rev. Lett. \textbf{90}, 127901
\bibitem{nisk1} Niskanen A O, Nakamura Y and Tsai J-S 1006 Phys. Rev. B {\bf 73}, 094506
\bibitem{niskanen} Niskanen A O, Harrabi K, Yoshihara F, Nakamura Y, Lloyd S, Tsai J S 2007 Science \textbf{316}, 723

%kwantowa optyka - Kerr nonlinearities
\bibitem{gerry} Gerry C C and Campos R A 2001 Phys. Rev. A {\bf 64}, 063814 %max enta photonic states
\bibitem{leonski} Leo\`{n}ski W and Miranowicz A 2004 J. Opt. B {\bf 6}, S37 %nlin coupler
\bibitem{yi} Yi X X, Zhou L, and Song H S 2004 J. Phys. A: Math. Gen. {\bf 37}, 5477 %two cavities entanglement through a nonlinear interaction
\bibitem{zhang} Zhang Z-M, Yang J, and Yu Y 2008 J. Phys. B: At. Mol. Opt. Phys. {\bf 41} 025502 %ent states of two states of light - Kerr interaction of laser beams
\bibitem{ottaviani} Ottaviani C, Vitali D, Artoni M, Cataliotti F, and Tombesi P 2003 Phys. Rev. Lett. {\bf 90},197902  %polarization qubit phase gate with kerr nonlinearity
\bibitem{vitali} Vitali D, Fortunato M, and Tombesi P 2000 Phys. Rev. Lett. {\bf 85}, %teleportacja z wykorzystaniem kerr nlin

\bibitem{gardiner} Gadiner C W, Zoller P 2000 {\it Quantum noise}, (Berlin: Springer) 
\bibitem{neg} Vidal G, Werner R F 2002 Phys. Rev. A {\bf 65}, 032314
\bibitem{vourdas1} Vourdas A 2006 J. Phys. A \textbf{39} R65-R141
\bibitem{perina} Perina J, Hradil Z, Jurco B 1994 {\it Quantum optics and fundamentals of physics}, (New York: Springer-Verlag)
\bibitem{ve} Everitt  M J, Clark T D, Stiffell P B, Vourdas A, Ralph J F, Prance R J, Prance H 2004 Phys. Rev. A \textbf{69}, 043804
\bibitem{makh} Makhlin Y, Sh\"{o}n G, and Shnirman A 1999 Nature {\bf 398}, 305
\bibitem{cos2} Jaynes E T, Cummings F W 1963 Proc.IEEE51, 89
\bibitem{peres} Peres A 1996 Phys. Rev. Lett. {\bf 77}, 1413
\bibitem{stef} Steffen M, Ansmann M, Bialczak R C, Katz N, Lucero E, McDermott R, Neeley M, Weig E M, Cleland A N, Martinis J M 2006 Science {\bf 313}, 5792, 1423
 \bibitem{alicki} Alicki R, Lendi K 1987 {\it Quantum dynamical semigroups
 and applications}, ({\it Lecture Notes in Physics} {\bf 286}), (Berlin: Springer)



 \end{thebibliography}
 \end{document}